\documentclass[10pt]{article}

\usepackage{graphicx,amsfonts}
\newtheorem{thm}{T{\sc HEOREM}}[section]
\newtheorem{lemma}[thm]{LEMMA}

\def\bp{\noindent{\it Proof. }}
\def\ep{\noindent{\hfill \fbox{}}}
\def\be{\begin{eqnarray}}
\def\ee{\end{eqnarray}}
\def\ds{\displaystyle}
\def\nn{\nonumber}
\def\le{\lefteqn}

\newcommand{\mapright}[1]{%
   \smash{\mathop{%
   \hbox to 1cm{\rightarrowfill}}\limits^{#1}}}
\newcommand{\mapleft}[1]{%
   \smash{\mathop{%
   \hbox to 1cm{\leftarrowfill}}\limits^{#1}}}
\newcommand{\maplleft}[2]{%
   \smash{\mathop{%
   \hbox to 1cm{\leftarrowfill}}\limits_{#1}^{#2}}}

\begin{document}

\title{The ILW hierarchy}
\author{Y Tutiya and J Satsuma}

\date{}
\maketitle
\begin{center}
{Graduate School of Mathematical
Sciences\footnote{TEL:+81-3-5465-7001,FAX:+81-3-5465-7012\\
EMAIL:tutiya@poisson.ms.u-tokyo.ac.jp\, \,satsuma@ms.u-tokyo.ac.jp},
University of
Tokyo, Komaba 3-8-1, Meguro-ku, Tokyo 153-8914, Japan}\\

\end{center}

\begin{abstract}
In this paper, we present a new hierarchy which includes the intermediate
long wave (ILW) equation at the lowest order.\ This hierarchy is thought to
be a novel reduction of the 1st modified KP type hierarchy.\ The framework
of our investigation is Sato theory.\
\end{abstract}


\section{Introduction}
\ \ It is well known that Sato theory was established by M.Sato around 1980
to give a unified viewpoint for integrable soliton equations [1].\ A lot of
important studies have been based on this magnificent theory since then.\ In
this paper, we focus on one of the basic ideas of the theory, summarized as
follows [2].\

``Start from  an ordinary differential equation and suppose that the
solutions satisfy certain dispersion relations.\ Then, as conditions the
coefficients must satisfy, we obtain a set of nonlinear partial differential
equations. If we assume a particular set of linear dispersion relations, we
obtain the KP hierarchy as the corresponding PDEs.''

It is known that most nonlinear partial differential equations which have
$N$-soliton solutions correspond to certain equations of KP hierarchy.\ But,
in the case of the intermediate long wave (ILW) equation, although its
$N$-soliton solutions and an inverse scattering transform were well-known
[3,4,5], the correspondence to the KP hierarchy has remained unclear.\

The ILW equation was proposed by Joseph [6] and  Kubota et al. [7] to
describe long internal gravity waves in a stratified fluid with finite
depth.\ It is written in the form
\be
u_t + \frac{1}{\delta }u_x + 2uu_x + T(u_{xx}) = 0 ,
\ee
where $T(\cdot )$ is the singular integral operator given by
\be
\ds T(u) = P\int ^\infty _{-\infty} \frac{1}{2\delta }\cot
\left[\frac{\pi}{2\delta }(\xi - x )\right] u(\xi) d\xi
\ee
($P$ represents the principal value of the integral).\ Depending on the
parameter $\delta$ (which controls the depth of the internal wave layer) (1)
reduces to the Korteweg-de Vries (KdV) equation as $\delta \to 0$ or to the
Benjamin-Ono (BO) equation as $\delta \to \infty $.\

In this paper, we present a new hierarchy which includes the ILW equation at
the lowest order.\ This hierarchy is thought to be a novel reduction of the
1st modified KP type hierarchy.\ In section 2, we propose a set of
differential-difference dispersion relations, and introduce a corresponding
1st modified KP type hierarchy.\ In section 3, this $2+1$dimensional
hierarchy will be reduced to a $1+1$dimensional hierarchy which contains the
ILW equation at its lowest order.\ In section 4, we discuss more general
dispersion relations of differential-difference type.\

\section{The dispersion relations and their corresponding hierarchy}
Let us introduce a pseudo-differential operator,
\be
W = 1 - \frac{U}{2}\partial^{-1} + w_2\partial^{-2} + w_3\partial^{-3} +
w_4\partial^{-4} +\cdots + w_m\partial^{-m},
\ee
where $U$ and $w_j\, (j = 1,2,\cdots )$ are functions of continuous
variables $t = (t_1,t_2,\cdots)$ and a discrete variable $z$.\ We sometimes
use $x$ instead of $t_1$ to follow the convention.\ $\partial^{-n}$ denotes
\be
\partial^{-n} = \left( \frac{d}{dt_1} \right)^{-n}.
\ee
If we employ the Leibniz rule,
\be
\partial^nf(t_1) = \sum_{r=0}^{\infty} \frac{n(n-1)\cdots
(n-r+1)}{r!}\left(\frac{\partial^rf}{\partial t_1^r}\right)\partial^{n-r},
\ee
then $\partial^n$ can be a well-defined operator even for negative integer
$n$.
Though the theory is developed for the case of $m \to \infty$ in general,
we, in this paper, confine ourselves to (3) for simplicity [2]. It is
remarked that the essence of the general theory is still kept in this
simplification.

Let us consider the ordinary differential equation,
\be
W\partial^m f(t,z) = (\partial^m - \frac{U}{2}\partial ^{m-1} + w_2\partial
^{m-2} +\cdots + w_m)f(t,z) = 0
\ee
which has $m$ linearly independent solutions,
$f^{(1)}(t,z),f^{(2)}(t,z),\cdots ,f^{(m)}(t,z)$.\ We assume here that
$f^{(1)}(t,z),f^{(2)}(t,z),\cdots ,f^{(m)}(t,z)$ satisfy the following
dispersion relations,
\be
\left\{ \begin{array}{l}
\ds i^{1-n}\partial _{t_n} f^{(j)} = \partial ^nf^{(j)}  \\
\ds \Delta_z f^{(j)}   = \partial f^{(j)}
       \end{array}
 \right. (j = 1,2,\cdots ,m\; \mbox{and} \; n = 1,2,\cdots),
\ee
where $\Delta_z$ denotes a difference operator,
\be
\begin{array}{lll}
\ds \Delta_z g(z)&=&\ds \frac{e^{2i\delta \partial _z} - 1}{2i\delta } g(z)
\\
 &=&\ds \frac{g(z + 2i\delta) - g(z)}{2i\delta }
\end{array}
\ee
($\delta$ is an arbitrary constant).\

It should be remarked that $U$ and $w_j$ are expressible by means of a
$\tau$-function, which is the Wronskian of $f^{(1)}(t,z),f^{(2)}(t,z),\cdots
,f^{(m)}(t,z)$, i.e.
\be
U = 2\frac{\tau_x}{\tau} \,,\,w_2 =
\frac{1}{2}\left(\frac{\tau_{t_2}}{\tau} -
\frac{\tau_{xx}}{\tau}\right)\,,\cdots .
\ee
Let $B_n \, (n = 1,2,\cdots)$ and $C$ be pseudo-differential operators,
\be
&&\ds B_n = (W\partial ^nW^{-1})_+, \\
&&\ds C = (\bar{W}\partial W^{-1})_+,
\ee
where $\bar{W}(z) = W(z + 2i\delta)$ and $(A)_+$ denotes the differential
part of the pseudo differential operator $A$.\ We use $\bar{\cdot}$ to
denote a shift operator: $z \to z + 2i\delta$ .\ Then we introduce time
evolution equations for W by
\be
\left\{
\begin{array}{l}
\ds i^{1-n}\frac{\partial W}{\partial t_n}= B_nW - W\partial ^n \\
\ds \frac{\bar{W} - W}{2i\delta } = CW - \bar{W}\partial
\end{array}
\right.
(n = 1,2,\cdots ),
\ee
which are Sato type equations.\ From (12) we get an infinite system of the
Zakharov-Shabat type equations
\be
\left\{
\begin{array}{l}
\ds i^{1-l}(C)_{t_l} - \frac{\bar{B}_l-B_l}{2i\delta } + CB_l - \bar{B}_lC =
0  \\
\ds i^{1-l}(B_k)_{t_l} - i^{1-k}(B_l)_{t_k} + [B_k , B_l] = 0\; \; \; \; \;
\; \; \; \;
\end{array}
\right.  \\
\hspace{7.5cm}(k,l = 1,2,\cdots ). \nn
\ee
Furthermore from (13) we can deduce a system of partial
differential-difference equations for $U$.\ The first few are explicitly
given by
\be
\le{ i(\bar{U} - U)_{t_2} + \frac{i}{\delta }(\bar{U} - U)_x + (\bar{U} -
U)(\bar{U} - U)_x + (\bar{U} + U)_{xx} = 0, }& & \\
\le{2(\bar{U} - U)_{t_3} - \frac{3i}{2}(\bar{U} - U)_{xt_2} +
\frac{1}{2}(\bar{U} - U)_{xxx}} & & \nn  \\
 & &+\, \frac{3i}{2\delta }(\bar{U} + U)_{xx} + \frac{6i}{2\delta}(\bar{U} -
U)(\bar{U} - U)_x - \frac{3}{2\delta_1 ^2}(\bar{U} - U)_x   \\
 & & +\, \frac{3}{2}(\bar{U} + U)_{xx}(\bar{U} - U) + \frac{3}{2}(\bar{U} +
U)_x(\bar{U} - U)_x + \frac{3}{2}(\bar{U} - U)^2(\bar{U} - U)_x = 0, \nn \\
\lefteqn{-i(\bar{U} - U)_{t_4} -\frac{1}{2} (\bar{U} + U)_{t_2t_2} +
\frac{i}{2}(\bar{U} - U)_{xxt_2} + (\bar{U} + U)_{xxxx} }& & \\
 & &+\, 3(\bar{U} - U)_x(\bar{U} - U)_{xx} - \frac{i}{2}(\bar{U} +
U)_{t_2}(\bar{U} - U)_x - i(\bar{U} - U)_{t_2}(\bar{U} + U)_x \nn \\
 & &-\, \frac{i}{2}(\bar{U} + U)_{xt_2}(\bar{U} - U) +
\frac{i}{2\delta^2}(\bar{U} - U)_{t_2} + \frac{1}{2\delta}(\bar{U} +
U)_{xt_2} + \frac{i}{\delta}(\bar{U} - U)_{xxx} \nn \\
& &+\, \frac{1}{\delta}(\bar{U} - U)(\bar{U} - U)_{t_2} + (\bar{U} -
U)(\bar{U} - U)_{xxx} - \frac{i}{2}(\bar{U} - U)^2(\bar{U} - U)_{t_2} = 0.\
\nn
\ee
Substituting (9) into (14)-(16), we also have the equations for $\tau $ by
\be
&&\left( iD_{t_2} + \frac{i}{\delta }D_x + D_x^2\right) \bar{\tau } \cdot
\tau  = 0, \\
&&\left(4D_{t_3} - 3iD_xD_{t_2} + D^3_x + \frac{3i}{\delta }D^2_x -
\frac{3}{\delta^2}D_x\right)\bar{\tau }\cdot \tau = 0, \\
&&\left(-2iD_{t_4} - D^2_{t_2} - iD_{t_2}D^2_x + \frac{1}{\delta }D_{t_2}D_x
+ \frac{i}{\delta^2}D_{t_2}\right) \bar{\tau }\cdot \tau = 0
\ee
($D$ denotes Hirota's differential operator).\ It should be remarked that
(17)-(19) are essentially the same as the first few of the 1st modified KP
hierarchy [8],
\be
&&(D_{t_1}^2 + D_{t_2})\tau_n \cdot \tau_{n+1} = 0, \\
&&(D_{t_1}^3 - 4D_{t_3} - 3D_{t_1}D_{t_2})\tau_n \cdot \tau_{n+1} = 0, \\
&&(-D_{t_1}^2D_{t_2} + D_{t_2}^2 + 2D_{t_4})\tau_n \cdot \tau_{n+1} = 0.
\ee
\section{Solutions and special reductions}
$N$-soliton solutions for (17)-(19) will be written in the form
\be
\tau &=&\left| \begin{array}{ccc}
1 + c_1e^{\eta(t,p_1) - \eta(t,q_1)}&\cdots&1 + c_Ne^{\eta(t,p_N) +
\eta(t,q_N)}  \\
l(q_1) + l(p_1)c_1e^{\eta(t,p_1) - \eta(t,q_1)}&\cdots&l(q_N) +
l(p_N)c_Ne^{\eta(t,p_N) - \eta(t,q_N)}  \\
\vdots&\cdots&\vdots \\
l(q_1)^{N-1} + l(p_1)^{N-1}c_1e^{\eta(t,p_1) -
\eta(t,q_1)}&\cdots&l(q_N)^{N-1} + l(p_N)^{N-1}c_Ne^{\eta(t,p_N) -
\eta(t,q_N)}
                \end{array}
             \right| \nn \\
& & /  \prod_{j'>j}(l(q_{j'}) - l(q_j)) \nn \\
&=& \sum_{J \subset I}\left(\prod_{j \in
J}c_j\right)\left(\prod_{\stackrel{j,j'\in
J}{j<j'}}a_{jj'}\right)\mbox{exp}\left(\sum_{j\in
J}\eta(t,p_j)-\eta(t,q_j)\right),
\ee
where the summation is taken over all subsets $J$ of $I = \{1,2,\cdots
,m\}$.\ $\eta(t,p) $, $l(p)$ and $a_{jj'}$ are defined by
\be
&&\eta(t,p) = pz + \sum_{n=1}^{\infty}i^{n-1}l(p)^nt_n,  \\
&&l(p) = \frac{e^{2i\delta p} - 1}{2i\delta }, \\
&&a_{jj'} = \frac{l(p_j) - l(p_{j'})}{l(p_j) - l(q_{j'})}\cdot
\frac{l(q_j) - l(q_{j'})}{l(q_j) - l(p_{j'})}\; ,
\ee
and the $c_j$ are constants.\ Notably, the bilinear identity to $\tau $ is
given as follows.\
\begin{lemma} For arbitrary $t = (t_1,t_2,\cdots) , t' = (t'_1,t'_2,\cdots)$
and $z$, $\tau$ satisfies
\be
\oint \mbox{e}^{-i\xi(k,t-t')}(1 + 2\delta
k)\tau(z+2i\delta,t-i\epsilon(k^{-1}))\tau(z,t+i\epsilon(k^{-1}))\frac{dk}{2
\pi i} = 0,\
\ee
where $\xi(k,t) = \sum_{n=0}^{\infty}t_nk^n\ ,\ \epsilon(k^{-1}) =
(\frac{1}{k},\frac{1}{2k^2},\cdots,\frac{1}{nk^n},\cdots)$.\ The curve is
taken around $\infty$ and excludes the singular points $il(p),il(q)$.\
\end{lemma}
\bp
Substitute (23) into (27) and we see that $Res(il(p_n)) + Res(il(q_n)) = 0$
for $\forall n$.\
\ep

We impose the condition
\be
p_j - q_j = l(p_j) - l(q_j)  = k_j
\ee
on (23).\ Then, it is reduced to
\be
\tau = \sum_{J \subset I}\left(\prod_{j \in
J}c_j\right)\left(\prod_{\stackrel{j,j'\in
J}{j<j'}}a_{jj'}\right)\mbox{exp}\left(\sum_{j\in J}\left(k_jz +
\sum_{n=1}^{\infty}\mu_n(k_j)t_n\right)\right)
\ee
for appropriately-defined functions $\mu_n(k)$.\ It should be noticed that
(29) is the same as the soliton solution of the ILW equation [9].\ We here
present the 1-soliton solution as an example.\
\be
\le{ \tau = 1+\ds c\,{\rm exp}\left[ kz + kt_1 + (k^2\cot k\delta -
\frac{k}{\delta })t_2\right. } \nn \\
 & & +\, \frac{1}{4}\left(k^3 - 3k^3\cot ^2k\delta +
\frac{6}{\delta }k^2\cot k\delta - \frac{3}{\delta ^2}k\right)t_3  \\
& &\left.+\, \frac{1}{2}\left(-k^4\cot k\delta  + k^4\cot ^3k\delta  -
\frac{3}{\delta }k^3\cot ^2k\delta  + \frac{1}{\delta }k^3 + \frac{3}{\delta
^2}k^2\cot k\delta - \frac{1}{\delta ^3}k\right)t_4 + \cdots \right] \nn
\ee
We can regard $z$ and $t_1$ as the same variable under this reduction,
because the coefficient of $z$ is equal to that of $t_1$ in the
exponentiated part of (29).\ It also should be noticed that
\begin{lemma}
If $k_j$'s are real, $\mu_n(k_j)$ and $a_{jj'}$ are also real $(j,n  =
1,2,\cdots )$.\
\end{lemma}
\bp

\noindent From (25) and (28), we get
\be
l(p_j) + l(q_j) = -ik_j\cot k_j\delta + \frac{i}{\delta}.
\ee
Because $l(p_j) + l(q_j)$ is purely imaginary and $l(p_j) - l(q_j) (= k_j)$
is real, there exist real $r,\theta$ by which
\be
l(p_j) = re^{i\theta}\; , \; l(q_j) = re^{i(\pi - \theta)}.
\ee
By the definition of $\mu_n(k_j)$, we have
\be
\mu_n(k_j) &=& i^{n-1}\left(l(p_j)^n - l(q_j)^n \right) \nn \\
            &=& i^{n-1}\left(re^{in\theta} - re^{in(\pi - \theta)}\right)
\nn \\
     &=& \left\{ \begin{array}{l}
2i^{n-1}r^n\cos(n\theta)\;\;\; \mbox{for}\; n\; \mbox{odd}  \\
2i^nr^n\sin(n\theta)\;\;\;\; \mbox{for}\; n\; \mbox{even}.
       \end{array}
 \right.
\ee
Hence $\mu_n(k_j)$ is real.\ We also deduce
\be
a_{jj'}&=& \frac{l(p_j) - l(p_{j'})}{l(p_j) - l(q_{j'})}\cdot \frac{l(q_j) -
l(q_{j'})}{l(q_j) - l(p_{j'})}\nn\\
 &=& \frac{(re^{i\theta} - r'e^{i\theta'})(re^{-i\theta} -
r'e^{-i\theta'})}{(re^{i\theta} - r'e^{i\theta'})(re^{i\theta} -
r'e^{i\theta'})}\nn\\
 &=&\frac{r^2 + r'^2 - rr'\cos(\theta - \theta')}{r^2 + r'^2 -
rr'\cos(\theta + \theta')}\; ,
\ee
which gives that $a_{jj'}$ is real.\
\ep

\
By this reduction, the $x$-shifts can take the place of the $z$-shifts and
(14)-(16) are rewritten into the equations for $U^\pm(x) := U(x\mp
i\delta)$, i.e.
\be
\le{i(U^- - U^+)_{t_2} + \frac{i}{\delta }(U^- - U^+)_x + (U^- - U^+)(U^- -
U^+)_x + (U^- + U^+)_{xx} = 0, }& & \\
 & & \nn \\
\le{2(U^- - U^+)_{t_3} - \frac{3i}{2}(U^- - U^+)_{xt_2} + \frac{1}{2}(U^- -
U^+)_{xxx} + \frac{3i}{2\delta }(U^- + U^+)_{xx} } & &   \\
 & & +\, \frac{3i}{\delta}(U^- - U^+)(U^- - U^+)_x -
\frac{3}{2\delta^2}(U^- - U^+)_x + \frac{3}{2}(U^- + U^+)_{xx}(U^- - U^+)
\nn \\
 & & +\, \frac{3}{2}(U^- + U^+)_x(U^- - U^+)_x + \frac{3}{2}(U^- -
U^+)^2(U^- - U^+)_x = 0, \nn \\
 & & \nn \\
\lefteqn{-i(U^- - U^+)_{t_4} -\frac{1}{2} (U^- + U^+)_{t_2t_2} +
\frac{i}{2}(U^- - U^+)_{xxt_2} + (U^- + U^+)_{xxxx} }& & \\
 & &+\, 3(U^- - U^+)_x(U^- - U^+)_{xx} - \frac{i}{2}(U^- + U^+)_{t_2}(U^- -
U^+)_x - i(U^- - U^+)_{t_2}(U^- + U^+)_x \nn \\
 & &-\, \frac{i}{2}(U^- + U^+)_{xt_2}(U^- - U^+) + \frac{i}{2\delta^2}(U^- -
U^+)_{t_2} + \frac{1}{2\delta}(U^- + U^+)_{xt_2} + \frac{i}{\delta}(U^- -
U^+)_{xxx} \nn \\
& &+\, \frac{1}{\delta}(U^- - U^+)(U^- - U^+)_{t_2} + (U^- - U^+)(U^- -
U^+)_{xxx} - \frac{i}{2}(U^- - U^+)^2(U^- - U^+)_{t_2} = 0. \nn
\ee

If we consider lemma 3.2 and suppose that $U(x)$ is analytic in the
horizontal strip between $\mbox{Im}\ x = -i\delta$ and $\mbox{Im}\ x =
i\delta$, we can introduce a dependent variable $u$ which satisfies [9]
\be
&&u = \frac{i}{2}(U^- - U^+), \\
&&T(u) = \frac{1}{2}(U^- + U^+).
\ee
Substituting this $u$ into (35)-(37), we obtain
\be
\le{ u_{t_2} + \frac{1}{\delta}u_x + 2uu_x + T(u_{xx}) = 0, }& & \\
 & & \nn \\
\le{ -4u_{t_3} - 3T(u_{xt_2}) - u_{xxx} - 6uT(u_{xx}) - 6u_xT(u_x) +
12u^2u_x }& &  \\
 & &\ds  - \frac{12}{\delta }uu_x + \frac{3}{\delta }T(u_{xx}) +
\frac{3}{\delta ^2}u_x = 0, \nn \\
 & & \nn \\
\lefteqn{-2u_{t_4} - T(u_{t_2t_2}) + u_{xxt_2} - 4u^2u_{t_2} -
2u_xT(u_{t_2}) - 4u_{t_2}T(u_x)}& &  \\
 & &\ds -2uT(u_{xt_2}) + 2T(u_{xxxx}) - 12u_xu_{xx} - 4uu_{xxx} \nn \\
 & &\ds + \frac{1}{\delta }T(u_{xt_2}) + \frac{2}{\delta }u_{xxx} -
\frac{4}{\delta }uu_{t_2} + \frac{1}{\delta ^2}u_{t_2} = 0. \nn
\ee
Because the lowest order is nothing but the intermediate long wave equation,
this hierarchy should be called the ILW hierarchy.\

\section{More general dispersion relations}
Other possible differential-difference dispersion relations than (7) can be
\be
\left\{ \begin{array}{l}
\ds i^{1-n}\partial _{t_n} f^{(j)} = \partial ^nf^{(j)} \\
\ds i^{s-1}\Delta _{z} f^{(j)} = \partial ^sf^{(j)}, \\
       \end{array}
 \right.
\ee
where $s$ is some fixed positive integer, $j = 1,2,\cdots ,m$ and $n =
1,2,\cdots$.\ Notably, (43) corresponds to (7) when $s = 1$.\

For each $s$, we can deduce the corresponding hierarchy in the same way as
in the preceding sections.\ Hence we just present the bilinear identity
which generates the hierarchy for $\tau$ [8],
\be
\oint \mbox{e}^{-i\xi(k,t-t')}(1 - 2\delta
(-k)^s)\tau(z+2i\delta,t-i\epsilon(k^{-1}))\tau(z,t+i\epsilon(k^{-1}))\frac{
dk}{2\pi i} = 0.\
\ee
The $N$-soliton solution to the hierarchy is written in the form
\be
\tau &=& \left| \begin{array}{ccc}
1 + c_1e^{H(t,p_1) - H(t,q_1)}&\cdots&1 + c_Ne^{H(t,p_N) + H(t,q_N)}  \\
L(q_1) + L(p_1)c_1e^{H(t,p_1) - H(t,q_1)}&\cdots&L(q_N) +
L(p_N)c_Ne^{H(t,p_N) - H(t,q_N)}  \\
\vdots&\cdots&\vdots \\
L(q_1)^{N-1} + L(p_1)^{N-1}c_1e^{H(t,p_1) - H(t,q_1)}&\cdots&L(q_N)^{N-1} +
L(p_N)^{N-1}c_Ne^{H(t,p_N) - H(t,q_N)}
                \end{array}
             \right| \nn\\
& &/  \prod_{j'>j}(L(q_{j'}) - L(q_j)) \nn \\
&= & \sum_{J \subset I}\left(\prod_{j \in
J}C_j\right)\left(\prod_{\stackrel{j,j'\in
J}{j<j'}}A_{jj'}\right)\mbox{exp}\left(\sum_{j\in
J}H(t,p_j)-H(t,q_j)\right),
\ee
where the summation is taken over all subsets $J$ of $I = \{1,2,\cdots
,m\}$.\ $H(t,p)$,$L(p)$ and $A_{jj'}$ are defined by
\be
&&H(t,p) = pz + \sum_{n=1}^{\infty}i^{n-1}L(p)^nt_n, \\
&&i^{1-s}L(p)^s = \frac{e^{2i\delta p} - 1}{2i\delta }, \\
&&A_{jj'} = \frac{L(p_j) - L(p_{j'})}{L(p_j) - L(q_{j'})}\cdot
\frac{L(q_j) - L(q_{j'})}{L(q_j) - L(p_{j'})}\; .\
\ee
It should be noticed that $L(p)$ is multi-valued.\ It is easy to check that
this soliton solution satisfies the biliniear identity (44).\

We impose the reduction condition,
\be
p_j - q_j &=& i^{s-1}\left( L(p_j)^s - L(q_j)^s\right)\nn \\
 &=&(-1)^{s-1}\frac{e^{2i\delta p_j} - e^{2i\delta q_j}}{2i\delta } \nn \\
 &=&k_j \hspace{3cm}(j = 1,2,\cdots).\
\ee
Substituting (49) into (45), we have
\be
\tau = \sum_{J \subset I}\left(\prod_{j \in
J}C_j\right)\left(\prod_{\stackrel{j,j'\in
J}{j<j'}}A_{jj'}\right)\mbox{exp}\left(\sum_{j\in J}\left(k_jz +
\sum_{n=1}^{\infty}M_n(k_j)t_n\right)\right)
\ee
for appropriately-defined functions $M_n(k)$.\ We can regard $z$ and $t_s$
as the same variable under this reduction.\ Furthermore, the following lemma
holds.\
\begin{lemma}
If $k_j$'s are real, $M_n(k_j)$ and $A_{jj'}$ are also real ($i =
1,2,\cdots$).\
\end{lemma}
\bp
>From (48) and (50), we see that
\be
&&L(p_j)^s - L(q_j)^s  = (i)^{1-s}k_j, \\
&&L(p_j)^s + L(q_j)^s  = (i)^{s}\left((-1)^sk_j\cot k_j\delta +
\frac{2}{\delta} \right).\
\ee
If $s$ is odd, $L(p_j)^s - L(q_j)^s$ is real and $L(p_j)^s + L(q_j)^s$ is
purely imaginary.\ Hence there exists real $r,0\leq \theta \leq 2\pi$ by
which
\be
L(p_j)^s = re^{i\theta}\; , \; L(q_j)^s = re^{i(\pi - \theta)}.
\ee
If $s$ is even, $L(p_j)^s - L(q_j)^s$ is purely imaginary and $L(p_j)^s +
L(q_j)^s$ is real.\ Then we have
\be
L(p_j)^s = re^{i\theta}\; , \; L(q_j)^s = re^{-i\theta}.\
\ee
For both cases, we choose $L(p_j), L(q_j)$ as
\be
L(p_j) = \sqrt[s]{r}e^{i\frac{\theta}{s}},\; L(q_j) = \sqrt[s]{r}e^{i(\pi -
\frac{\theta}{s})}.\
\ee
Because $L(p_j) - L(q_j)$ is defined to be real and $L(p_j) + L(q_j)$ to be
purely imagenary, we see that $M_j, A_{jj'}$ are real by means of lemma
3.2.\ This, at the same time, takes care of the multi-valuedness of the
function $L(p)$.\
 \ep

If we consider lemma 4.1 and suppose that $U(t_s)$ is analytic in the
horizontal strip between $\mbox{Im}\ t_s = -i\delta$ and $\mbox{Im}\ t_s =
i\delta$, we can introduce a dependent variable u which satisfies
\be
&&u = \frac{i}{2}(U(t_s + i\delta) - U(t_s - i\delta)), \\
&&T_s(u) = \frac{1}{2}(U(t_s + i\delta) + U(t_s - i\delta)),
\ee
where
\be
\ds T_s(u(t_s)) = P\int ^\infty _{-\infty} \frac{1}{2\delta }\cot
\left[\frac{\pi}{2\delta }(\xi - t_s )\right] u(\xi) d\xi .\
\ee

As another concrete example, different from the ILW hierarchy, we apply the
preceding argument to the case $s  = 2$.\ From (44), we get
\be
\oint \mbox{e}^{-i\xi(k,t-t')}(1 - 2\delta
k^2)\tau(z+2i\delta,t-i\epsilon(k^{-1}))\tau(z,t+i\epsilon(k^{-1}))\frac{dk}
{2\pi i} = 0,
\ee
which generates the hierarchy for $\tau$.\ The first few equations are
\be
&&\left(-2D_{t_3} + D^3_x + 3iD_xD_{t_2} +
\frac{3}{\delta}D_x\right)\bar{\tau }\cdot \tau = 0, \\
&&\left(6D_{t_4} -4iD_{t_3}D_{x}-3iD_{t_2}^2 -
iD_{x}^4 -\frac{12i}{\delta}D_{x}^2 -
\frac{6}{\delta}D_{t_2}\right)\bar{\tau }\cdot \tau = 0.
\ee
This hierarchy is essentially the same as the 2nd modified KP hierarchy [8]
\be
&&\left(2D_{t_3} + D^3_x + 3D_xD_{t_2}\right) \tau_n \cdot \tau_{n+2} = 0,
\\
&&\left(D_1^4 - 4D_1D_3 - 3D_2^2 - 6D_4\right) \tau_n \cdot \tau_{n+2} = 0,
\\
&&\cdots .\nn
\ee
The $1$-soliton solution for this hierarchy is written in the form
\be
&&\tau = 1 + C{\rm exp}[H(t,p) - H(t,q)],\\
&&H(t,p) = (p-q)z + \sum_{n=1}^{\infty}i^{n-1}L(p)^nt_n,\\
&&L(p)^2 = \frac{-e^{2i\delta p} + 1}{2\delta }.\
\ee
By the reduction condition
\be
p - q &=& i\left( L(p)^2 - L(q)^2\right)\nn \\
 &=&-\frac{e^{2i\delta p} - e^{2i\delta q}}{2i\delta } \nn \\
 &=&k,
\ee
(64) is reduced to
\be
\le{\tau =  1 + {\rm exp}\left[ kz + \sqrt{ k\cot k\delta - \delta^{-1} +
\sqrt{(k\cot k\delta - \delta^{-1})^2 + k^2}  }\: t_1 + kt_2 \right. }& & \\
 & &\ds +\, \sqrt{ k\cot k\delta - \delta^{-1} + \sqrt{(k\cot k\delta -
\delta^{-1})^2 + k^2}  }\nn \\
 & &\hspace{2cm}\left. \times \, \left(  -k\cot k\delta + \delta^{-1} +
\sqrt{(k\cot k\delta - \delta^{-1})^2 + k^2}/\sqrt{2} \right)t_3 + \cdots
\right].\ \nn
\ee
Thus the $t_2$-shifts take the place of the $z$-shifts.\ If we impose the
analytic condition as before on $U$, there exists $u$ which satisfies
\be
&&u = \frac{i}{2}(U(t_2 + i\delta) - U(t_2 - i\delta)), \\
&&T_2(u) = \frac{1}{2}(U(t_2 + i\delta) + U(t_2 - i\delta)).\
\ee
Now, (60) and (61) are expressible by means of $u$ as
\be
\le{ 2u_{t_3} - u_{xxx} + 3T_2(u_{xt_2}) - 3\left(uT_2(u_x)\right)_x +
3u^2u_x }& &  \\
& &\ds -\, 3\left(u\int_{-\infty}^x u_{t_2} dx\right)_x -
\frac{3}{\delta}u_x = 0, \nn \\
\le{ 6u_{t_4} + T_2(u_{xxxx}) - 4(uu_x)_x + 12T_2(u)T_2(u)_x + 4u^3u_x }& &
\\
& &-6\left(6u^2T_2(u_x)  \right)_x + 4T_2(u_{xt_3}) -
4u_x\int_{-\infty}^{x}u_{t_3}d\xi - 4uu_{t_3} + 3T_2(u_{t_2t_2})  \nn \\
& &- 6u_{t_2}\int_{-\infty}^{x}u_{t_2}d\xi + \frac{12}{\delta}T_2(u_{xx}) -
\frac{24}{\delta}uu_x - \frac{6}{\delta}u_{t_2} = 0.\ \nn
\ee
{\noindent{\bf Acknowledgment.}}
This work was partially supported by a Grant-in-Aid from the Japan
foundation for the Promotion of Science (JSPS).\ The authors would like to
thank T. Tokihiro, R. Willox and Y. Ohta for discussions and advice.\


\end{document}